\documentclass[11pt,a4paper]{article}
\pdfoutput=1
\usepackage{jcappub}

\usepackage{graphicx}
\usepackage{amsmath}
\usepackage{epsfig,multicol,bbm}
\usepackage{subcaption}

\begin{document}

\title{\boldmath How rare is the Bullet Cluster\\ (in a $\Lambda$CDM universe)?}

\abstract{The Bullet Cluster (1E\,0657-56) is well-known as providing
  visual evidence of dark matter but it is potentially incompatible
  with the standard $\Lambda$CDM cosmology due to the high relative
  velocity of the two colliding clusters. Previous studies have
  focussed on the probability of such a high relative velocity amongst
  selected candidate systems. This notion of `probability' is however
  difficult to interpret and can lead to paradoxical results. Instead,
  we consider the expected number of Bullet-like systems on the sky up
  to a specified redshift, which allows for direct comparison with
  observations. Using a Hubble volume N-body simulation with high
  resolution we investigate how the number of such systems depends on
  the masses of the halo pairs, their separation, and collisional
  angle. This enables us to extract an approximate formula for the
  expected number of halo-halo collisions given specific collisional
  parameters. We use extreme value statistics to analyse the tail of
  the pairwise velocity distribution and demonstrate that it is fatter
  than the previously assumed Gaussian form. We estimate that the
  number of dark matter halo pairs as or more extreme than 1E\,0657-56
  in mass, separation and relative velocity is $1.3^{+2.0}_{-0.6}$ up
  to redshift $z=0.3$. However requiring the halos to have collided
  and passed through each other as is observed decreases this number
  to only 0.1. The discovery of more such systems would thus indeed
  present a challenge to the standard cosmology.}

\author[a]{David Kraljic}
\author[a,b]{\& Subir Sarkar}

\affiliation[a]{Rudolf Peierls Centre for Theoretical Physics,
  University of Oxford,\\ 1 Keble Road, Oxford, OX1 3NP, United
  Kingdom}
\affiliation[b]{Niels Bohr International Academy, University of
  Copenhagen,\\ Blegdamsvej 17, 2100 Copenhagen, Denmark}
\emailAdd{David.Kraljic@physics.ox.ac.uk}\emailAdd{S.Sarkar@physics.ox.ac.uk}
\arxivnumber{1412.7719}
\maketitle
\flushbottom

\section{Introduction}

Large clusters of galaxies are the biggest gravitationally bound
objects in the universe. Na\"ively they would be expected to
correspond to `fundamental observers' who follow the Hubble flow and
move \emph{away} from each other. The discovery of the Bullet Cluster
(1E\,0657-56), a system consisting of two very massive clusters of
galaxies which have undergone a collision with a high relative
velocity, thus requires an assessment of whether the standard
$\Lambda$CDM cosmology is able to accommodate such an extreme
event. The subsequent discovery of many more merging
clusters\footnote{Listed on
  \url{http://www.mergingclustercollaboration.org/}} motivates a
general study of the statistics of such events. Since such collisions
involve non-linear interactions, the predictions of the $\Lambda$CDM
model have to be extracted from cosmological N-body simulations.

The Bullet Cluster located at $z=0.296$ is the best studied of such
mergers, since its collisional trajectory is normal to the line of
sight. It is extreme in several respects. The main cluster has a high
mass of $M_{\rm main}\simeq 1.5 \times 10^{15}M_\odot$, with the
subcluster mass $M_{\rm Bullet}\simeq 1.5 \times 10^{14}M_\odot$,
separation $d_{12}\simeq0.72$\,Mpc
\citep{clowe_weak-lensing_2004,bradac_strong_2006}, and a very high
velocity $v_{12}\simeq4500$\,km/s \citep{markevitch_direct_2004}
deduced from the analysis of the bow shock. This should be compared to
the expected separation velocity in the Hubble flow: $v_{\rm
  Hubble}=H_{z=0.3}\times d_{12}\simeq70$\,km/s, where $H_z \simeq
H_0(1+z)\sqrt{1+\Omega_m z}$ and $H_0 \equiv 100h$ \,km/s/Mpc is the
Hubble parameter at $z=0$ with $h \simeq 0.7$. However the relative
velocity of the two dark matter (DM) peaks does not necessarily
correspond to the shock front velocity of the baryons observed in
X-rays, and a lower estimate of $v_{12}\simeq3000$\,km/s was given in
Ref.\cite{farrar_new_2007}.

Recent hydrodynamical simulations indeed show that the morphology (DM
and gas) of the Bullet system is well reproduced by requiring $v_{12}
\simeq 3000$\,km/s and $d_{12} \simeq 2.5$\,Mpc/$h$ between the DM
halos at redshift $z=0.5$
\citep{lage_constrained_2014,lage_bullet_2014}. This is broadly
comparable to Ref.\citep{mastropietro_simulating_2008} which had
estimated earlier: $v_{12} \simeq 3200$\,km/s at $d_{12} \simeq
2.5$\,Mpc/$h$. Simpler hydrodynamical simulations done earlier
\citep{springel_speed_2007, milosavljevic_cluster_2007} had also found
a lower relative velocity to reproduce the morphology better. The
extreme properties of 1E\,0657-56 can now be framed in terms of these
initial conditions required to produce the observed collision.

Searching for Bullet-like systems satisfying these initial conditions
in N-Body simulations is convenient as the two clusters can be taken
to be well separated and are thus easily identified. However because
of its rarity the Bullet system can only be found in the largest
N-Body simulations, provided the mass resolution is good (i.e. the DM
`particle' mass is small). Ref.\cite{thompson_pairwise_2012} shows
that to find a Bullet-like system, the volume of the simulation needs
to be at least $\sim(4.5$\,Gpc/$h)^3$, with the mass resolution better
than $M_{\rm par} \sim 6.5\times 10^{11} M_{\odot}/h$.

Various definitions of what constitutes a Bullet-like system have been
employed. Ref.\cite{hayashi_how_2006} used a 0.5\,(Gpc/$h$)$^3$
simulation and looked at the most massive subclusters moving away from
the host cluster with velocity $>4500$\,km/s and a separation
$>0.7$\,Mpc/$h$ at redshift $z=0.3$. With such a small simulation
volume they did not find any host halo as massive as 1E\,0657-56 so
needed to extrapolate to find the fraction of appropriate
subclusters. They estimated this to be $\sim 0.01$, however this is
uncertain to at least an order of magnitude.

Ref.\cite{lee_bullet_2010} used a (3\,Gpc/$h)^3$ simulation with
$M_{\rm par}\sim 2.3\times 10^{11} M_{\odot}/h$ to look at the
fraction of subhalos with high enough velocity among systems
satisfying the conditions in
Ref.\cite{mastropietro_simulating_2008}. This fraction, dubbed
`probability', was found to be about $10^{-10}$ at $z=0$, by fitting a
Gaussian to the pairwise velocity distribution in order to estimate
the tail of the distribution. This was interpreted as an inconsistency
between $\Lambda$CDM and the observation of 1E\,0657-56.

Ref.\cite{thompson_pairwise_2012} studied the effect of the box size
and the resolution on the tail of the pairwise velocity distribution
and found that simulations with small boxes and poor resolution
struggle to produce systems as extreme as 1E\,0657-56. Using rather
different definitions for a Bullet-like system than
Refs.\cite{hayashi_how_2006} and \cite{lee_bullet_2010} they estimated
the fraction of systems with high enough relative velocity by fitting
a skewed Gaussian to the pairwise velocity distribution. Again dubbed
`probability', this was estimated to be $\sim 10^{-8}$, still very
inconsistent with $\Lambda$CDM.

Ref.\cite{bouillot_probing_2014} used a large simulation with volume
of (21\,Gpc/$h)^3$ but poor resolution $M_{\rm par}\sim 1.2\times
10^{12} M_{\odot}/h$. They also used a different percolation parameter
($b=0.15$ instead of the conventional $b=0.2$) in their Friend of
Friends (FoF) halo finder, arguing that this more faithfully
reconstructs the masses of halos corresponding to the Bullet
system. The tail of the pairwise velocity distribution was analysed in
the Extreme Value Statistics (EVS) approach to show that it is
significantly fatter than a Gaussian-like tail. Using a similar
definition for Bullet-like systems as
Ref.\cite{thompson_pairwise_2012} they found that the fraction (again
called `probability') of such high-velocity encounters is about $\sim
6 \times 10^{-6}$, again raising a problem for $\Lambda$CDM.

Ref.\cite{thompson_rise_2014} explored the effect of halo finders on
the pairwise velocity distribution. In particular, a
configuration-space based FoF algorithm was compared to
\textsc{rockstar} \cite{behroozi_rockstar_2013}, a phase-space based
halo finder. Using the same definition for a Bullet-like system as in
Ref.\cite{thompson_pairwise_2012}, it was found that the FoF halo
finder fails to identify the collisions in the high-velocity tail and
leads to `probabilities' almost two orders of magnitude lower that
when the better performing \textsc{rockstar} halo finder is used.

Finally Ref.\cite{watson_statistics_2014} used a (6\,Gpc/$h)^3$ volume
with $M_{\rm par}\sim 7.5\times 10^{11} M_{\odot}/h$ and argued that a
Bullet-like system \citep{springel_speed_2007} at $z=0.3$ is not too
far from other halo pairs in the simulation. Instead of focussing on
the extreme properties of colliding dark matter halos,
Ref.\cite{forero-romero_bullet_2010} looked at the morphology of the
gas and dark matter in the colliding clusters and found that the
displacement between the gas and dark matter similar to 1E\,0657-56 is
expected in about 1\% of the clusters. Some other investigations
\cite{farrar_new_2007,bouillot_probing_2014,baldi} have even
considered whether the apparent inconsistency posed by the Bullet
Cluster can be alleviated by invoking a new long range `fifth' force
in the `dark sector'.  Rather than engage in such speculations we
address in this paper the main shortcoming of the previous studies,
viz. the ill-defined `probability' of finding systems like 1E\,0657-56
on the sky.

We use Dark Sky Simulations \citep{skillman_dark_2014}, one of the
biggest N-Body simulations with volume (8\,Gpc/$h)^3$, as well as one
of the best resolutions $M_{\rm par}\sim 3.9\times 10^{10}
M_{\odot}/h$. The halo catalogue used was produced by a phase-space
based (\textsc{rockstar}) halo finder that performs better than the
configuration-space based finders used earlier. We carefully explore
the dependence of the pairwise velocity distribution on the different
definitions of Bullet-like systems. Furthermore, the machinery of EVS
is used to examine the tail of the distribution, rather than
\emph{assuming} that a Gaussian fit is a good description. We also
study an observationally better motivated quantity, viz. the absolute
number of bullet-like systems expected in a survey up to some
particular redshift. This should be contrasted with the fraction of
extreme objects (in a population of less extreme objects) that has
been the focus of earlier studies.

In Section \ref{sec:sims} we describe the N-Body simulation used and
demonstrate the importance of using a phase-space based halo finder
for searching for systems similar to 1E\,0657-56. Section
\ref{sec:EVS} summarises the EVS tools relevant for modelling the
tails of distributions. In Section \ref{sec:prob} we show how the
expected number of Bullet-like systems can be estimated. We also
discuss some of the paradoxical features of the `probability' ---
defined as a fraction of Bullet-like systems in a population of
candidate systems --- that has been studied previously. Section
\ref{sec:results} contains the main results of our paper, viz. the
expected number of systems similar to 1E\,0657-56 in $\Lambda$CDM.

\section{Simulations and halo finders}
\label{sec:sims}
The biggest dataset of the Dark Sky Simulation (DS) Early Data Release
\citep{skillman_dark_2014} has a box of volume (8\,Gpc/$h)^3$ with
$N^3=10240^3$ `dark matter particles' corresponding to $M_{\rm
  par}\sim 3.9\times 10^{10} M_{\odot}/h$. Such a large volume and
good resolution make it ideal for the study of rare objects like the
Bullet Cluster \citep{thompson_pairwise_2012}.

Halos in the DS simulation were identified with the \textsc{rockstar}
\citep{behroozi_rockstar_2013} halo finder. For computational
convenience we reduced the halo catalogue by requiring $M_{\rm halo}>
3.5 \times 10^{13} M_{\odot}/h$. \textsc{rockstar} is a phase-space
based halo finder and therefore performs better at identifying
Bullet-like systems which are characterised by a small distance
between the two massive clusters with a high relative
velocity. Standard Friend of Friend (FoF) halo finders work in
configuration space and therefore struggle to tell the two nearby
clusters apart --- this leads to \emph{underestimation} of the number
of Bullet-like systems in N-Body simulations. In phase space however
the host and the bullet clusters are well separated due to the high
relative velocity, hence can be correctly identified as two separate
clusters.

To illustrate this point, the host and the bullet halo are generated
at the DS simulation resolution, using the NFW
\citep{navarro_structure_1996} density profile as it best fits weak
lensing data on 1E\,0657-56 \citep{farrar_new_2007}. The DM particles
are given the velocities as in Ref.\cite{hernquist_n-body_1993}. The
two halos are placed at various distances with their relative velocity
set at 3000\,km/s. Then both \textsc{rockstar} and a FoF algorithm
with percolation parameter $b=0.2$ are used to extract the halo
information.  At separations of 5\,Mpc/$h$ and 4\,Mpc/$h$ both halo
finders identify the two halos correctly. However at 3\,Mpc/$h$ the
FoF algorithm identifies 2 halos only $\sim$30\% of the time, while at
2\,Mpc/$h$ and below it identifies only 1. By contrast,
\textsc{rockstar} finds 2 halos at all separations. Thus we expect a
depletion of the number of nearby mergers when a FoF based halo finder
is used, leading to underestimation of the number of objects similar
to the Bullet system.

\section{Extreme value statistics}
\label{sec:EVS}
Events in the tail of the pairwise velocity distribution need to be
modelled without assuming a functional form for the underlying
distribution and EVS provides a framework for doing so. We briefly
outline the formalism relevant to this study following
Ref.\cite{coles2001introduction}.

We are interested in modelling the statistical behaviour of extreme
values of a random variable $X$. The probability that $X$ exceeds a
specified high threshold $\mu$ is:

\begin{equation}
{\rm Pr}\left\lbrace X > \mu + x| X > \mu\right\rbrace = \frac{1-
  F(\mu+x)}{1-F(\mu)}
\label{eqn:probdef}
\end{equation}
Here $F(x)={\rm Pr}\left\lbrace X < x\right\rbrace$ is the cumulative
distribution function which is unknown and needs to be estimated.

The central result of EVS is that the maxima of a sequence of random
variables $X_{\rm max}=\left\lbrace X_1,...,X_N\right\rbrace $ with a
common cumulative distribution function $F(x)$ tend to be distributed
in the limit $N\rightarrow\infty$ according the Generalized Extreme
Value distribution $G(x)$:

\begin{equation}
\label{eqn:max}
{\rm Pr}\left\lbrace X_{\rm max}<x \right\rbrace=F_{X_1}(x) \times
... \times F_{X_N}(x)= F^{N}(x) \approx G(x)
\end{equation}
where
\begin{equation}
G_{\mu,\sigma,\xi}(x) = \exp \left\lbrace - \left[ 1+ \xi \left(
  \frac{x-\mu}{\sigma} \right) \right]^{-1/\xi} \right\rbrace
\label{eqn:G}
\end{equation}
for some $\mu, \sigma > 0$ and $\xi$.  From Eqs.(\ref{eqn:max}) and
(\ref{eqn:G}) we can estimate $F(x)$ and thus approximate
Eq.(\ref{eqn:probdef}) by the Generalized Pareto Distribution (GPD)
$H(x)$:
\begin{align}
&{\rm Pr}\left\lbrace X > \mu + x| X > \mu\right\rbrace = 1 -
  H_{\mu,\sigma,\xi}(x), \quad \mathrm{where:} &H_{\mu,\sigma,\xi}(x)
  = 1- \left[1+ \xi \left( \frac{x-\mu}{\sigma} \right)
    \right]^{-1/\xi}
\end{align}
with the condition $1+ \xi \left( (x-\mu)/\sigma \right) > 0$. The
above expression is fitted to the extreme events and provides a model
independent description of the tails of probability distributions. If
the underlying distribution is `Gaussian-like' (e.g. Gaussian or
skewed Gaussian), the tail parameter $\xi$ equals 0. Longer tails have
$\xi>0$ whereas shorter ones have $\xi<0$.

The next step in modelling the tail is the choice of the threshold
$\mu$. If it is too low, the asymptotically valid GPD does not apply
and our estimate will be biased. If $\mu$ is too high, the reduced
number of extreme events available results in high variance in the
estimated parameters. Provided the GPD description is valid above some
threshold $\mu_1$ then it is also valid for a higher threshold
$\mu_2>\mu_1$ with new parameters
($\mu_2,\sigma_{\mu_2},\xi_{\mu_2}$). However, the tail parameter
$\xi_\mu$ and the combination $\sigma_{\mu} - \mu \xi_{\mu}$ should
remain constant. Therefore the simplest method for the appropriate
choice of the threshold $\mu$ focusses on finding a region of
stability of these parameter combinations. In order to minimise the
variance, the \emph{lowest} $\mu$ consistent with stability is finally
chosen as the threshold.

\section{The number (versus `probability') of Bullet-like systems}
\label{sec:prob}
A Bullet-like system can be defined by cuts in the collisional
parameters describing the merger of two clusters. Such a definition is
particularly suited for the DM-only N-Body simulations. The
collisional parameters are the separation $d_{12}$ between the two
halos, the two masses $m_1$ and $m_2$, the relative speed $v_{12}$,
and the angle $\theta$ between the relative velocity and the
separation. To simplify the problem the mass cut is often made in
terms of the average mass $\langle M \rangle$ of the two halos and we
shall do so too.

The most prominent feature of 1E\,0657-56 is the high relative
velocity of its subcluster with respect to the main cluster. Thus both
the pairwise velocity distribution $\mathrm{d}n/\mathrm{d}v_{12}$ and
its cumulative $n(>v_{12})$ (where $n$ is the number density of the
halo pairs) will be studied.

The observationally relevant quantity is the expected number of
Bullet-like objects up to some redshift. However, the quantity studied
so far has been the fraction of such objects with respect to less
extreme candidate systems (i.e. having a lower relative velocity)
\citep{hayashi_how_2006,thompson_rise_2014,bouillot_probing_2014,thompson_pairwise_2012,lee_bullet_2010}. This
fraction is then interpreted as the `probability' of finding a
Bullet-like system although it is not directly related to the
likelihood of observing such an object on the sky. In terms of the
number densities it is expressed as $p_v=n(v_{12}>v_{\rm Bullet} \mid
\textrm{other cuts})/n(v_{12}> 0 \mid \textrm{other cuts})$. The
probability defined in this way is relative to the objects defined by
the initial mass, distance, and angle cuts and has a non-trivial and
sometimes paradoxical dependence on those cuts. For example,
increasing the mass cut in the definition of a Bullet-like system
leads to an increase in the `probability', even though the actual
number of systems has been \emph{reduced} drastically (see
Figs.~\ref{masscuts}-\ref{CDFmasscuts}).

Alternatively, the high masses of the two colliding clusters can be
taken as the main defining parameters and the `probability' written as
$p_m = n (m_1, m_2 > M_{\rm main}, M_{\rm Bullet} \mid \textrm{other
  cuts}) /n( m_1, m_2 > 0 \mid \textrm{other cuts})$, where the
relative velocity cut has now been taken before the mass cut. Even
though we are looking at the \emph{same} Bullet-like objects, one
finds $p_v \neq p_m $ simply due to the different order of the cuts
taken in the collisional parameters.

In what follows we focus therefore on the observationally motivated
and intuitively accessible quantity, viz. the expected number of
Bullet-like systems on the sky up to a specified redshift. This can be
expressed (in a flat universe) as:
\begin{equation}
N(<z) = \int_0^z n(v_{12} > v_{\rm Bullet}, \mathrm{cuts} \mid z') 4
\pi D_{\rm c}^2(z')\, \mathrm{d}D_{\rm c}(z'),
\label{eqn:number}
\end{equation}
where $n(v_{12}>v_{\rm Bullet}, \mathrm{cuts} \mid z)$ is the comoving
number density of Bullet-like objects at redshift $z$, and $D_{\rm
  c}(z)$ is the comoving distance to $z$.

Estimating the pairwise velocity function
$\mathrm{d}n/\mathrm{d}v_{12}$ and its cumulative version $n(>v_{12})$
in large simulations at many different redshifts can be
computationally expensive. However, $\mathrm{d}n/\mathrm{d} v_{12}$,
and consequently $n(>v_{12})$, were found to have a stable shape up to
$z\sim 0.5$ \citep{bouillot_probing_2014,thompson_pairwise_2012}. This
simplifies the analysis and we can approximate:
\begin{equation}
n(v_{12}>v_{\rm Bullet}, \mathrm{cuts} \mid z) \approx \alpha (z)
\times n(v_{12}>v_{\rm Bullet}, \mathrm{cuts} \mid z=0),
\label{eqn:approx}
\end{equation}
where the normalisation $\alpha (z)$ is proportional to the number of
pairs of halos satisfying specific cuts (mass, distance \ldots) and is
set equal to 1 at $z=0$. When one halo has a mass above $m_1$ and the
other above $m_2$, with their separation less than $d_{12}$, it can be
written as:
\begin{align}
\alpha (z) \propto \iiint \limits_{m_1, m_2, 0}^{~~~\infty, \infty,
  d_{12}} \frac{\mathrm{d}n_\mathrm{h}}{\mathrm{d}m_1'} \bigg |_z
\frac{\mathrm{d}n_\mathrm{h}}{\mathrm{d}m_2'} \bigg |_z \left[1 +
  \xi_\mathrm{hh}(r, m_1', m_2', z)\right] 4\pi r^2 \mathrm{d}r \,
\mathrm{d}m_2' \, \mathrm{d}m_1',
\label{eqn:alpha}
\end{align}
where $(\mathrm{d}n_\mathrm{h}/\mathrm{d}m )|_z$ is the halo mass
function at redshift $z$ and $\xi_\mathrm{hh}(r,m_1,m_2,z)$ is the
two-point correlation function of halos of mass $m_1$ and $m_2$ which
is conventionally expressed as $b (m_1, r,z) b (m_2, r,z) \xi_{\rm
  lin}(r, z)$ in terms of the halo bias $b$ (which includes the
non-linear correction). Non-trivial mass cuts are simply implemented
by including an appropriate window function in Eq.(\ref{eqn:alpha}).

\begin{figure}
\centering
\begin{subfigure}[b]{0.5\columnwidth}
\includegraphics[width=\columnwidth]{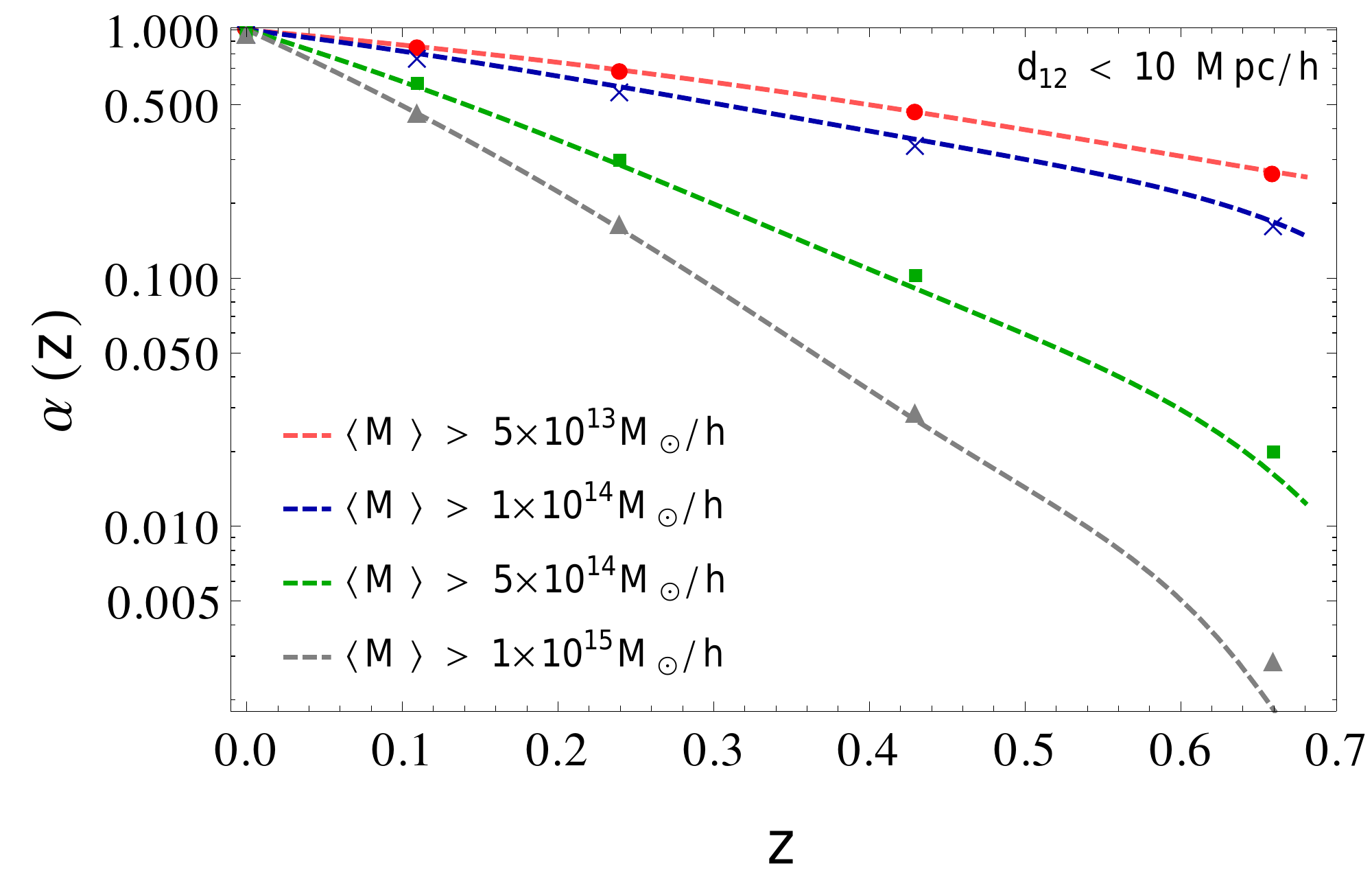}
\caption{The normalisation $\alpha(z)$ from Eq.(\ref{eqn:alpha}) for
  various mass cuts, compared to simulation `data'.}
\label{fig:alphamass}
\end{subfigure}%
~~
\begin{subfigure}[b]{0.5\columnwidth}
\includegraphics[width=\columnwidth]{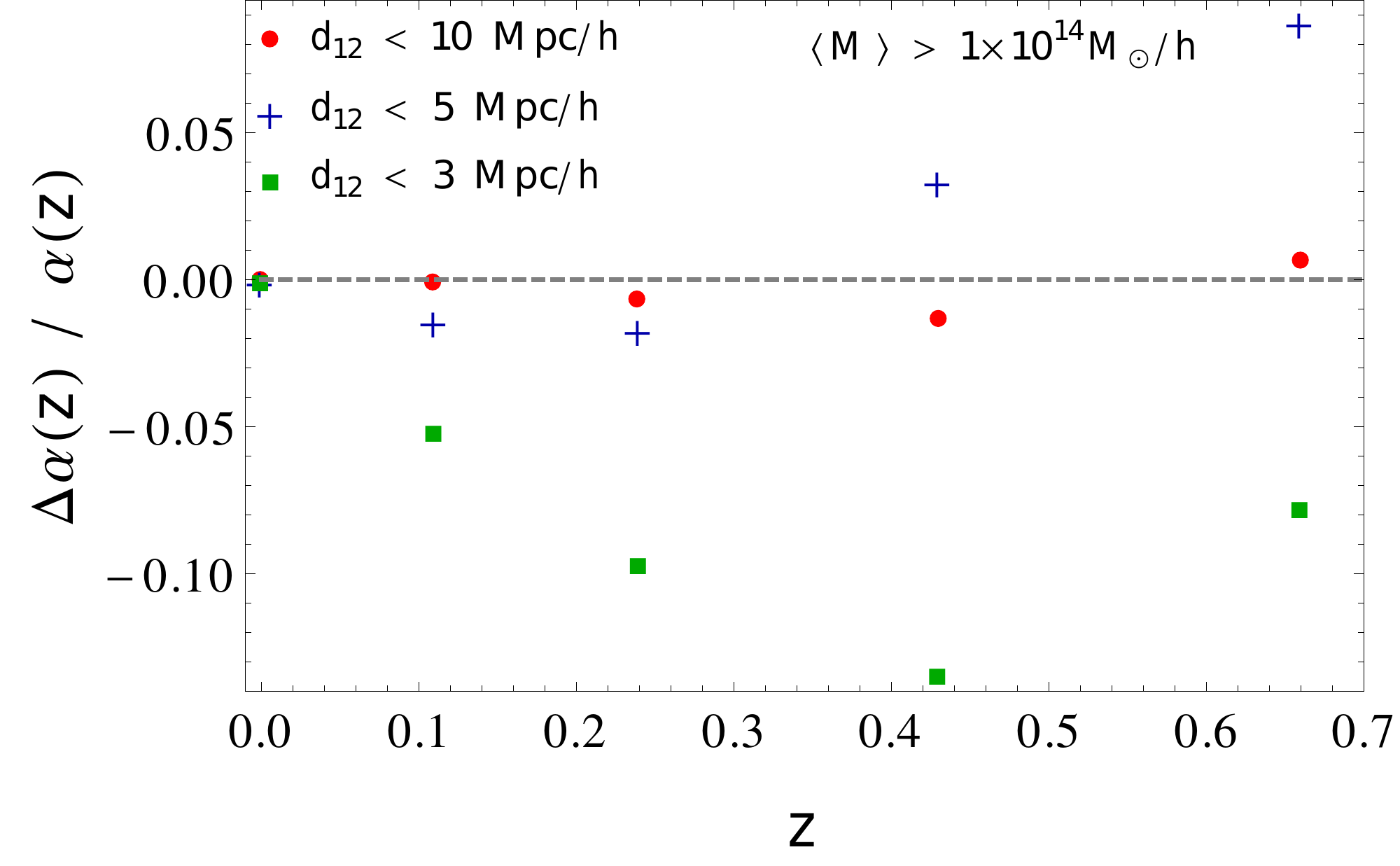}
\caption{The fractional difference in $\alpha(z)$ between the value
  extracted from the simulation and calculated using
  Eq.(\ref{eqn:alpha}), for different separation cuts.}
\label{fig:alphamass}
\end{subfigure}%

\begin{subfigure}[b]{0.5\columnwidth}
\includegraphics[width=\columnwidth]{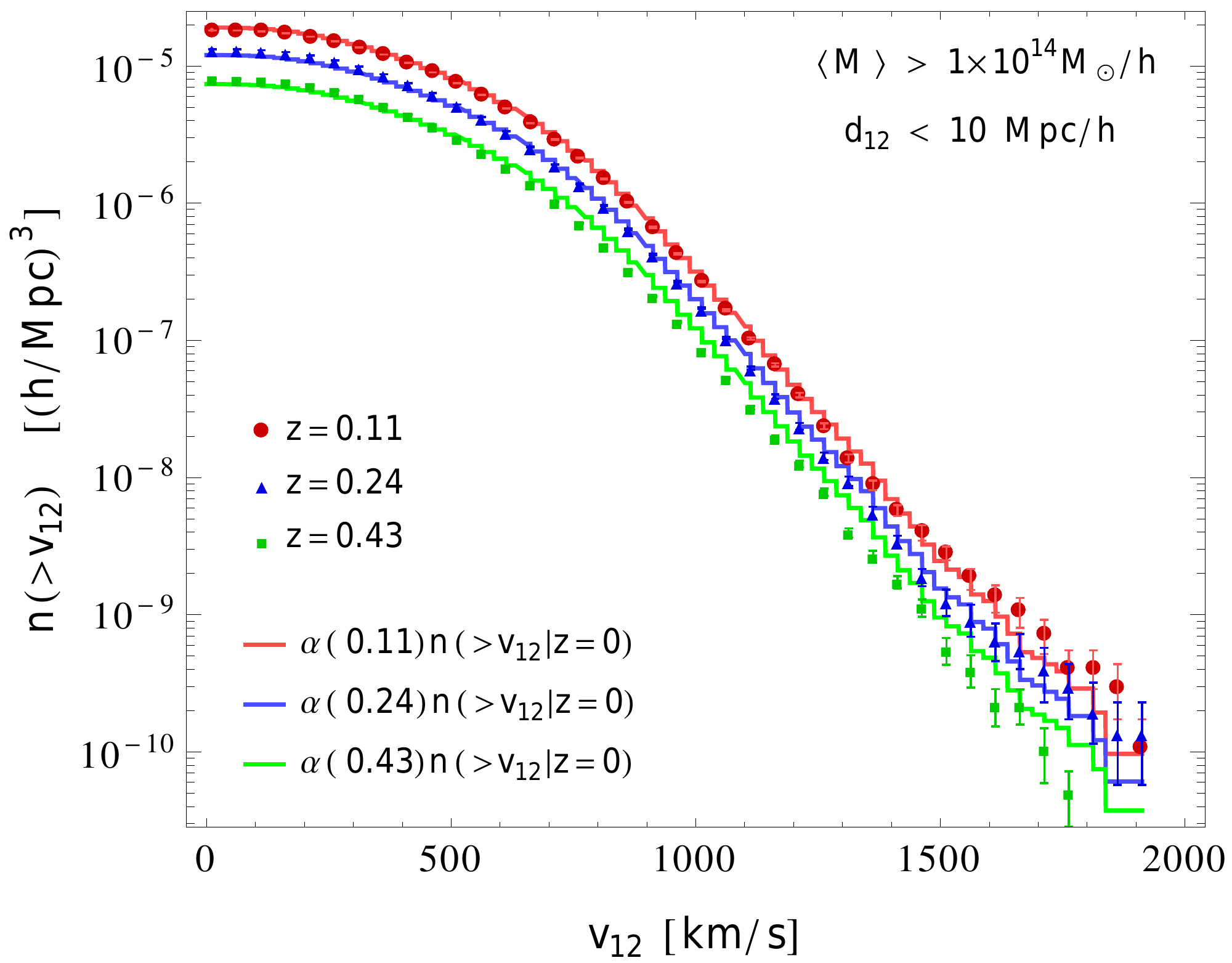}
\caption{Simulated `data' on cumulative pairwise velocity distribution
  at various redshifts, compared to the approximation
  (\ref{eqn:approx}). Errors shown are estimated by bootstrapping.}
\label{fig:nz}
\end{subfigure}
\caption{Testing our semi-analytic model (\ref{eqn:approx}--\ref{eqn:alpha})
  against the DEUSS-Lambda N-Body simulation.}
\label{alpha}
\end{figure}

Our semi-analytical expression (\ref{eqn:alpha}) provides an excellent
description of N-body simulations as illustrated in
Fig.\ref{alpha}. We use a DEUSS-Lambda \citep{rasera_introducing_2010}
N-Body simulation (containing 2048$^3$ particles in a
(2592\,Mpc/$h)^3$ volume, using a FoF halo finder) that is small
enough to be analysed at several redshifts. We have used the halo mass
function from Ref.\cite{tinker_large-scale_2010}, the power-spectrum
from CAMB (\url{http://camb.info}), and the best-fit halo bias formula
from Ref.\cite{tinker_mass--light_2005} in the expression
(\ref{eqn:alpha}). The normalisation $\alpha(z)$ is extracted by
taking the ratio $n(>v_{12}=0|z)/n(>v_{12}=0|z=0)$ which is then
compared to the value from
Eqs.(\ref{eqn:approx}-\ref{eqn:alpha}). Fig.\ref{alpha} shows that our
semi-analytic model becomes less accurate at high redshifts, high
masses and small distances as is expected. Bigger simulations are
required to explore these extreme regions in parameter space.

From Eqs.~(\ref{eqn:number}-\ref{eqn:alpha}) it then follows that the
number of Bullet-like systems factorises as:
\begin{equation}
N(<z) \approx n (v_{12} > v_{\rm Bullet}, \mathrm{cuts} \mid z=0)
\times V_{\rm eff}(z),
\end{equation}
where:
\begin{equation}
V_{\rm eff} (z) = \int_0^z \alpha(z') 4 \pi D_c^2 (z')\,\mathrm{d}
D_\mathrm{c} (z').
\label{V_eff}
\end{equation}
Therefore, the pairwise velocity distribution can be studied at $z=0$
in simulation outputs, provided we can estimate (either
semi-analytically as in Eq.(\ref{eqn:alpha}), or from a set of smaller
N-Body simulations) the effective volume $V_{\rm eff}$. This
simplification is valid in the observationally interesting redshift
range $z < 1$.

Given that $V_{\rm eff}$ is big enough (such that the clustering of
objects of interest is negligible), we can treat $N(<z)$ as being
Poisson distributed. Then the probability that we see at least one
object up to redshift $z$ is just: ${\rm Pr} \lbrace N \geq 1 \rbrace = 1 -
{\rm Poisson} \lbrace N=0 \mid \langle N \rangle = N(<z) \rbrace$.

\section{Results}
\label{sec:results}

Now we study the high-velocity tail of the pairwise velocity
distribution, in particular its dependence on the collisional
parameters --- the average mass, the relative distance, the
collisional angle, and the relative velocity of the halo pairs --- and
the correlations among these. The relative velocity of halo pairs,
$v_{12}$, is considered in \emph{proper} coordinates, i.e. including
the Hubble flow term $\mathbf{v}_{\rm H} = H\mathbf{d}_{12}$. Using
Eq.(\ref{eqn:approx}) we analyse the output of the N-Body simulation
(Sec.~\ref{sec:sims}) at redshift $z=0$.

\begin{figure}
\centering
 \includegraphics[width=0.5\columnwidth]{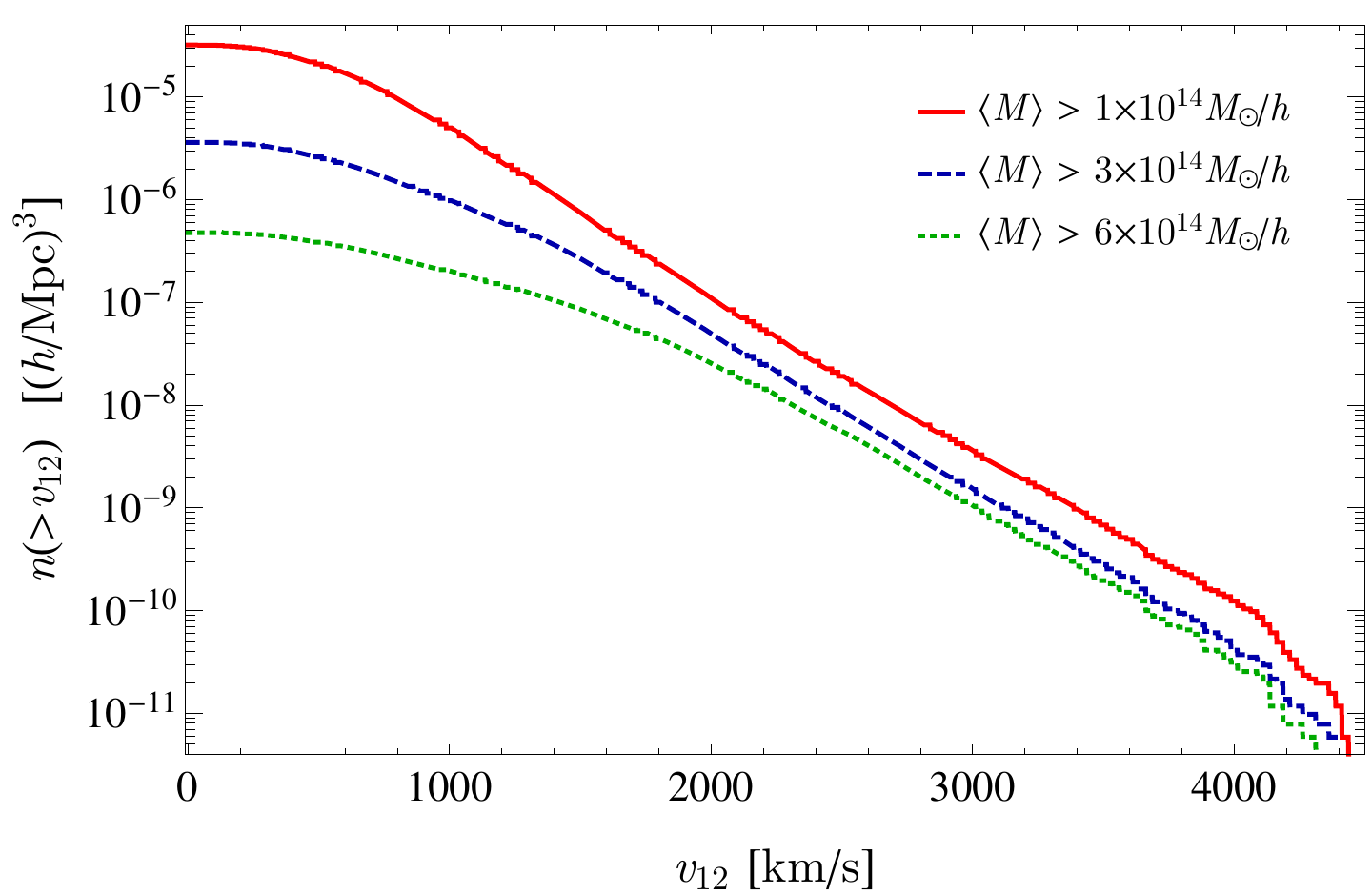}
 \caption{The dependence of the cumulative pairwise velocity
   distribution on the average mass $\langle M \rangle$ of the halo
   pairs. The halo mass function is steeply descending with $\langle M
   \rangle$, and so is the
   total number density of halo pairs, $n(>v_{12}=0|\langle M \rangle)$.}
   \label{masscuts}
\end{figure}

\begin{figure}
\centering
 \includegraphics[width=0.5\columnwidth]{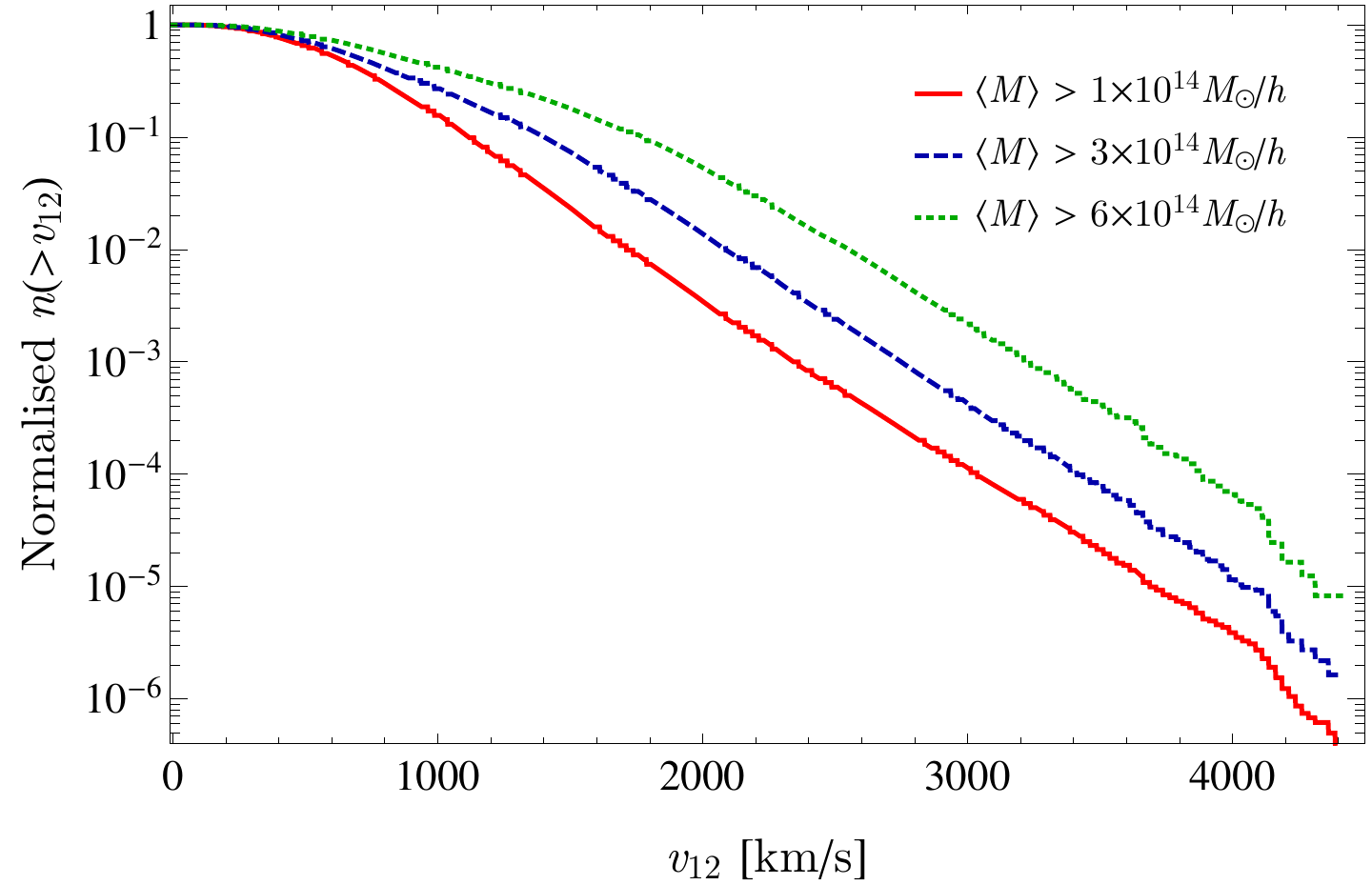}
 \caption{The cumulative pairwise velocity distribution normalised
   separately for each cut in the average mass.  In the set of
   more massive halos, the fraction of higher relative velocity pairs is
   bigger. Na\"ively this would be interpreted as a higher
   probability, however Fig.~\ref{masscuts} shows that the absolute
   number density of more massive halos pairs \emph{decreases} with
   $\langle M
   \rangle$.}
\label{CDFmasscuts}
\end{figure}

Increasing the cut in the average mass $\langle M \rangle$ of the halo
pairs, while keeping other collisional parameters (in particular
$v_{12}$) unchanged, the number density of the halo pairs
\emph{decreases} as seen in Fig.~\ref{masscuts}. By contrast, if we
chose to normalise the velocity distribution for each mass cut (as is
done in
Refs.\citep{hayashi_how_2006,thompson_rise_2014,bouillot_probing_2014,thompson_pairwise_2012,lee_bullet_2010}),
we would conclude that the `probability', i.e. the fraction of the
high-velocity collisions, increases (see Fig.~\ref{CDFmasscuts}).

\begin{figure}
\centering
 \includegraphics[width=0.5\columnwidth]{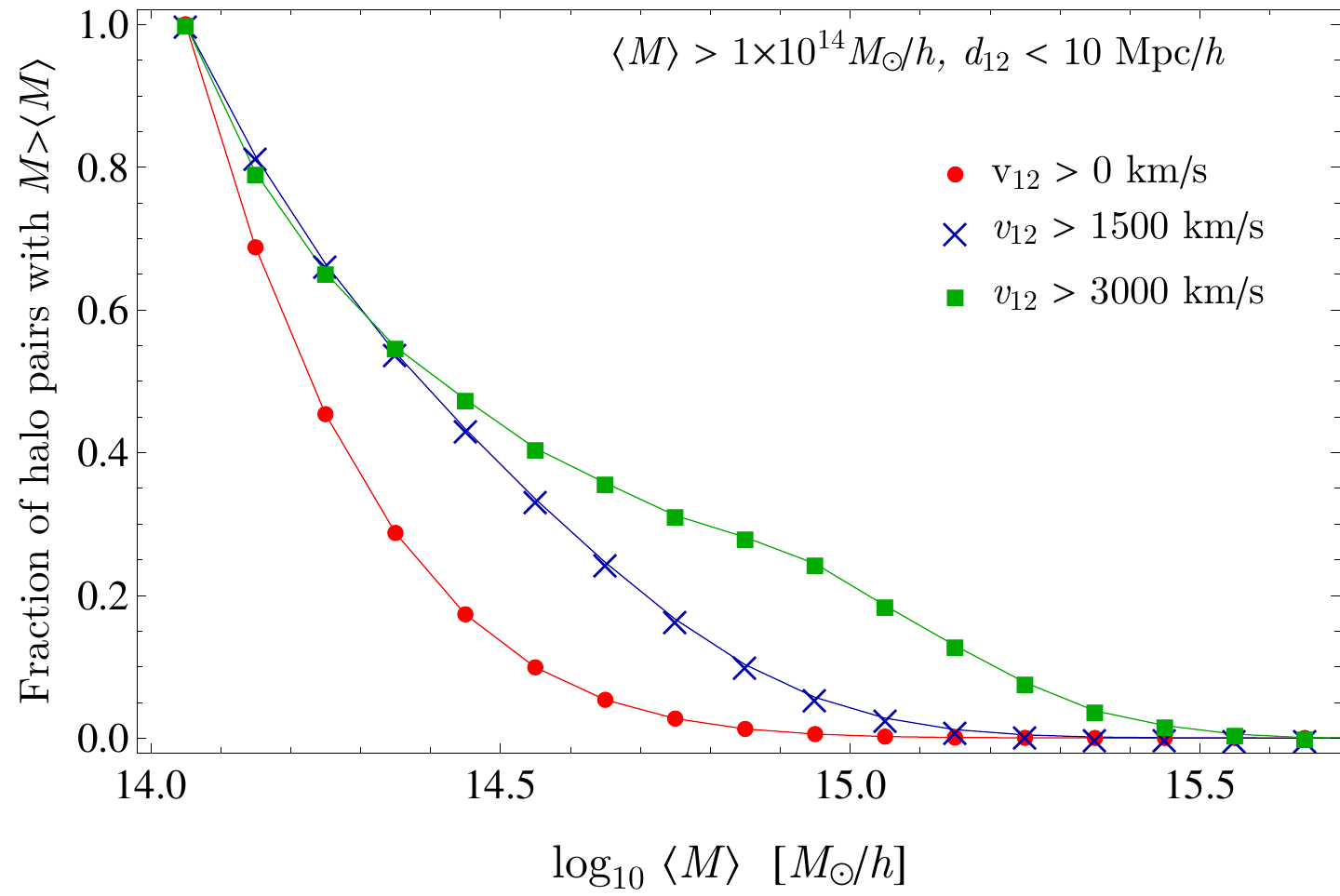}
\caption{The fraction of halo pairs above a specified average mass $\langle
  M \rangle$, given the relative velocity $v_{12}$. The fraction of
  high-mass halo pairs is \emph{higher} in the high-velocity
  population. Here $\langle M \rangle>10^{14} M_\odot /h$ and
  $d_{12}<10$\,Mpc/h.}
\label{massvelocity}
\end{figure}

In Newtonian gravity, for a bound system with mass $m$ we expect
$v_{12} \propto \sqrt{m}$. Therefore more massive halo pairs are
likely to have a higher relative velocity. Indeed in
Fig.~\ref{masscuts} the tail of the low-$\langle M \rangle$ velocity
distribution converges to the high-$\langle M \rangle$ tail at large
relative velocities, indicating that the tail of the pairwise velocity
distribution consists mostly of very massive halos. This is also seen
in the mass distribution of halo pairs in the high-velocity tail (see
Fig.~\ref{massvelocity}). Therefore, small N-Body simulations that
fail to produce halos with high masses \emph{underestimate} the tail
of the pairwise velocity distribution.

\begin{figure}
\centering
 \includegraphics[width=0.5\columnwidth]{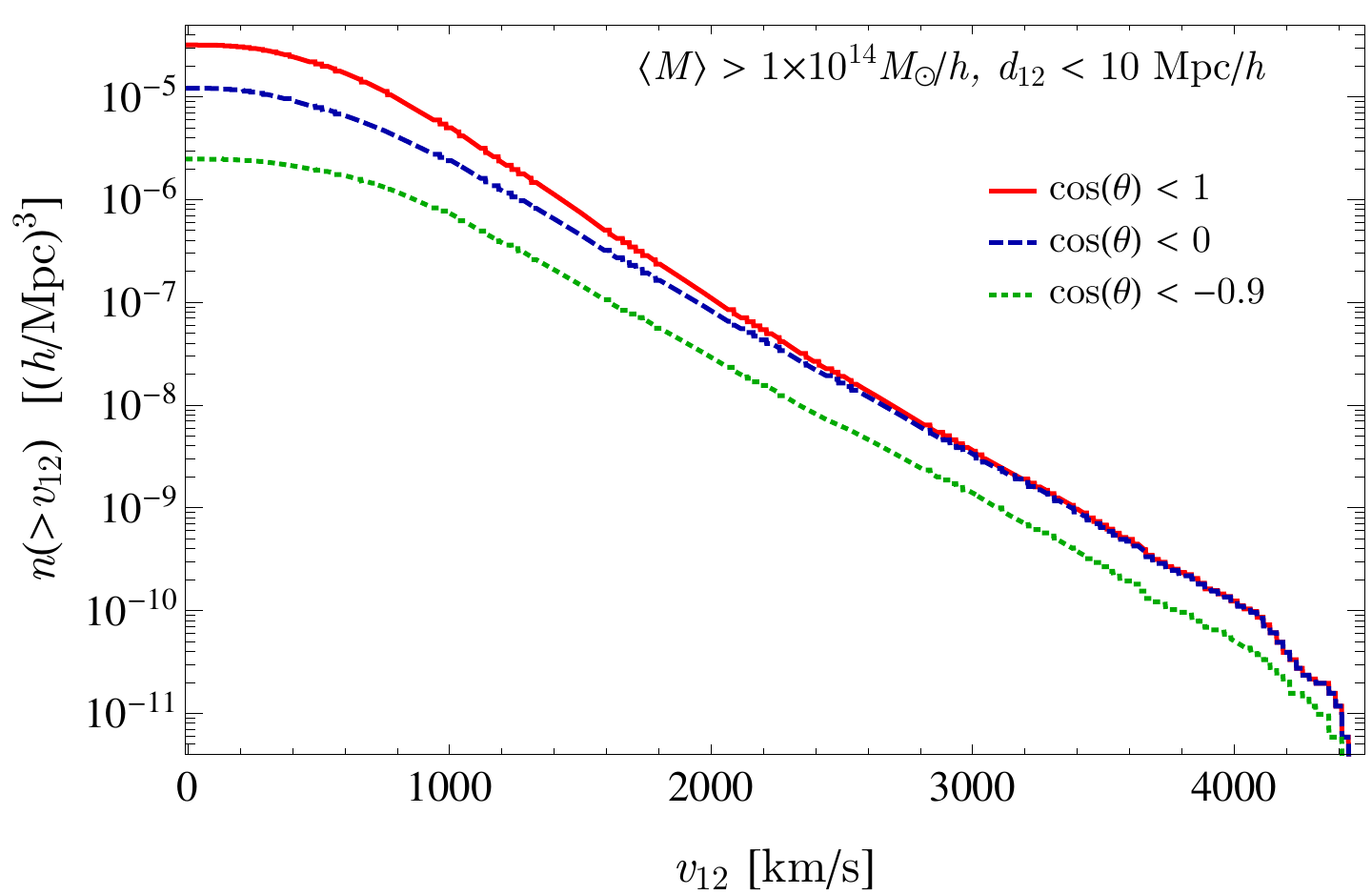}
\caption{Cumulative pairwise velocity distribution for different cuts
  on the collisional angle $\theta$, taking $\langle M \rangle>10^{14}
  M_\odot /h$ and $d_{12}<10$\,Mpc/h. Note that the high-velocity tail
  consists almost entirely of halos moving \emph{towards} each other
  ($\cos \theta < 0$).}
\label{anglecuts}
\end{figure}

\begin{figure}
\centering \includegraphics[width=0.5\columnwidth]{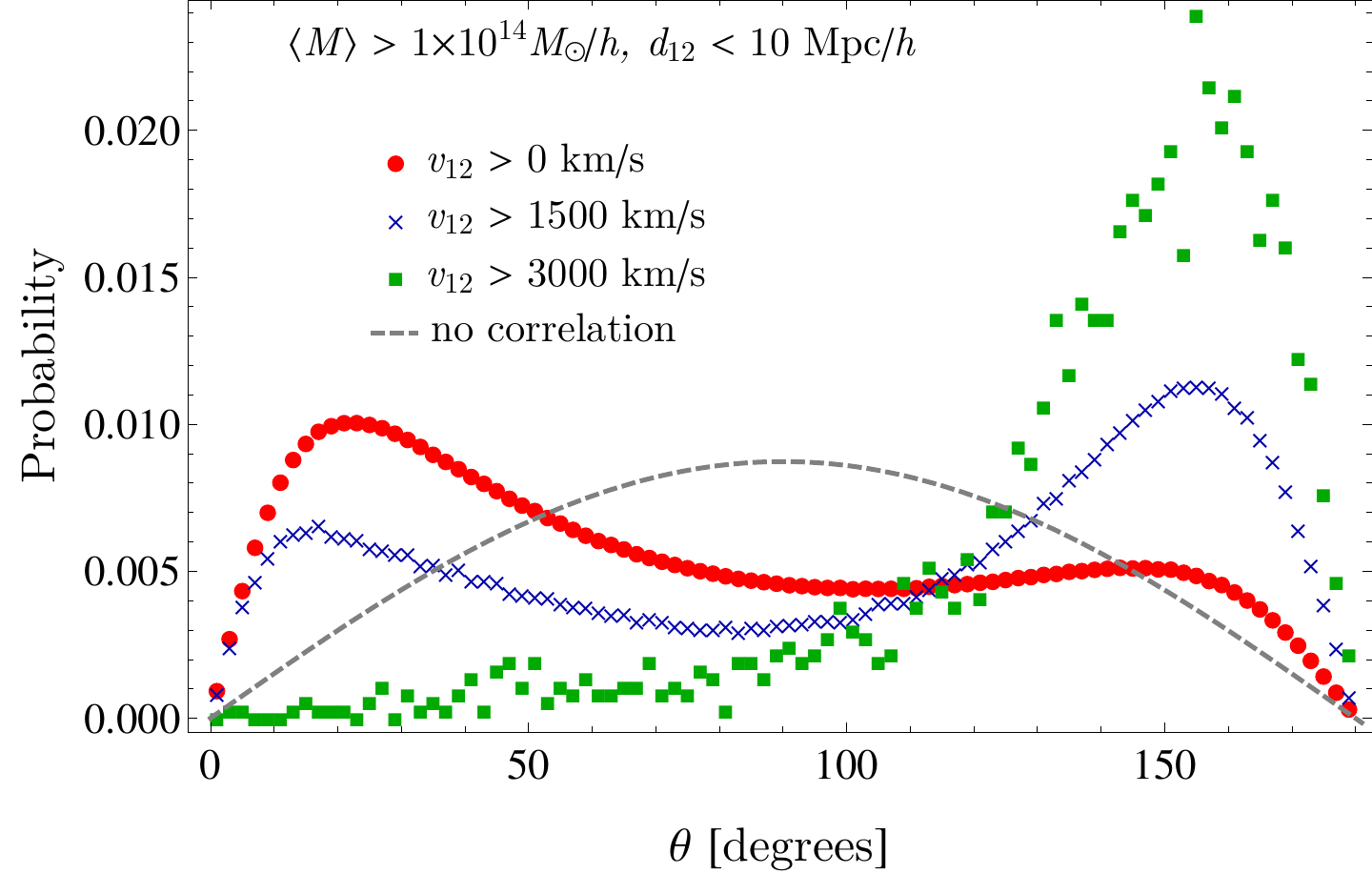}
 \caption{The probability distribution for the collisional angle
   $\theta$ given the relative velocity $v_{12}$, taking $\langle M
   \rangle>10^{14} M_\odot /h$ and $d_{12}<10$\,Mpc/h. Small relative
   velocities are mainly associated with the Hubble flow, $\theta \sim
   0^{\circ}$, whereas high velocities are mainly head-on collisions,
   $\theta \simeq 180^{\circ}$. The dashed line shows the case when the
   separation vector and the relative velocity vector are
   uncorrelated.}
\label{anglevelocity}
\end{figure}

The next collisional parameter we examine is the angle $\theta$
between the separation and the relative velocity of a halo pair. In
Fig.~\ref{anglecuts} we see that the tail of the velocity distribution
consists almost entirely of the halo pairs approaching each other
($\cos \theta < 0$). However, the number density of colliding halo
pairs is not as sensitive to cuts in the angle as to the cuts in the
halo masses. Again, as expected from Newtonian gravity, halo
collisions with high relative velocity are more likely to be
approximately head-on, as seen in Fig.~\ref{anglevelocity}.

\begin{figure}
\centering
 \includegraphics[width=0.5\columnwidth]{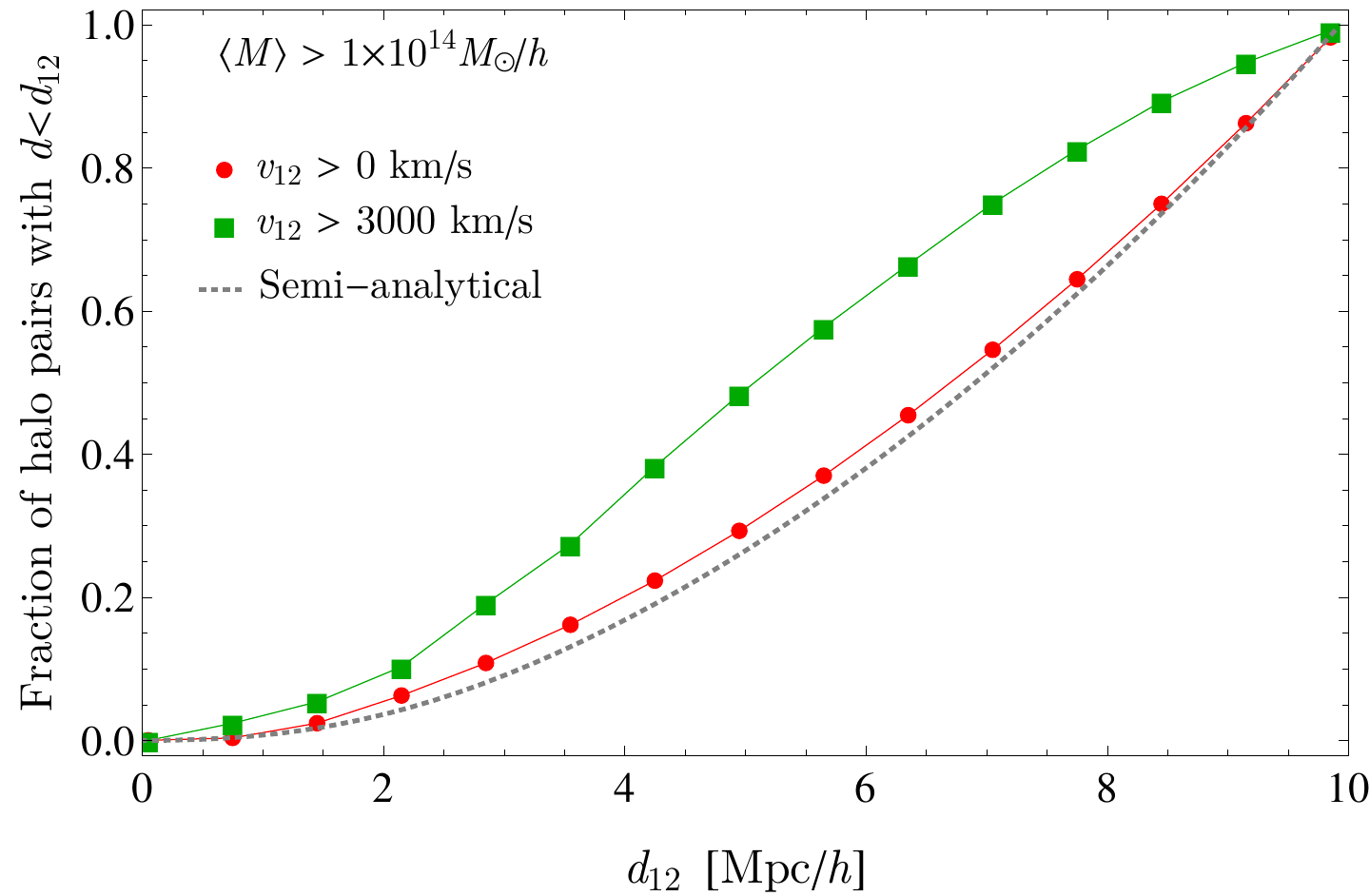}
 \caption{The fraction of halo pairs below a specified separation,
   given some relative velocity. For high-velocity collisions the
   halos are closer than for low pairwise velocities. The dashed line
   is the semi-analytical expectation for $v_{12}>0$\,km/s, calculated
   using Eq.(\ref{eqn:alpha}) with the cosmological parameters
   matching the Dark Sky simulation.}
\label{distancevelocity}
\end{figure}

In Fig.~\ref{distancevelocity} we see that the halo pairs with a high
relative velocity are more likely to be \emph{closer} together
compared to the low pairwise velocity. Therefore, a
configuration-space based halo finder (e.g. FoF) will miss relatively
more high velocity mergers compared to the low velocity ones, and
hence bias the tail of the pairwise velocity distribution to be
shorter. This characteristic of the halo finders has been explored in
greater detail in Ref.\cite{thompson_rise_2014}.

\begin{figure}
\centering
 \includegraphics[width=0.5\columnwidth]{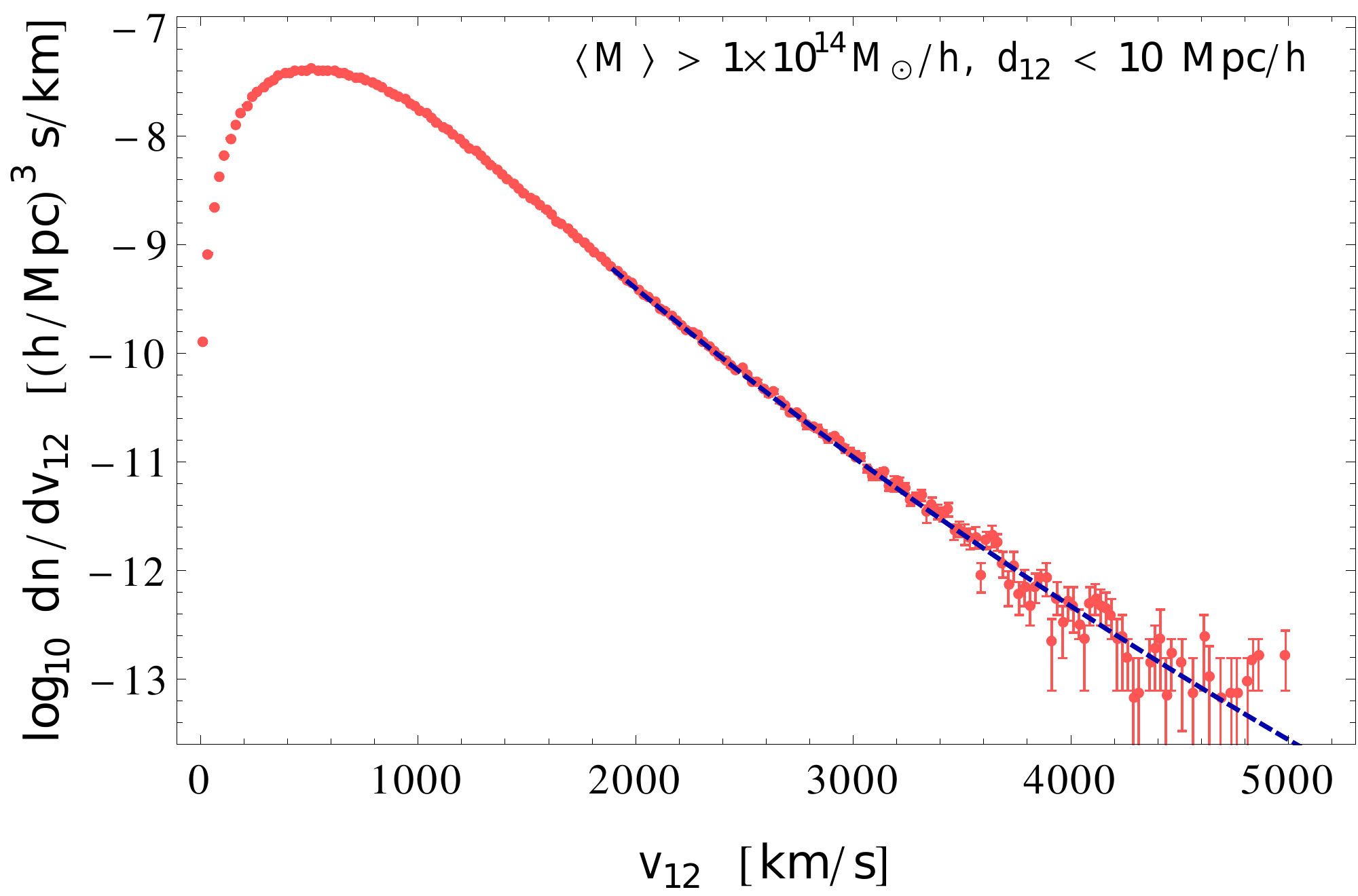}
\caption{Pairwise velocity distribution from the simulation at
  redshift $z=0$ compared to the best fit Generalised Pareto
  Distribution.}
\label{dndv12}
\end{figure}

\begin{figure}
\centering \includegraphics[width=0.5\columnwidth]{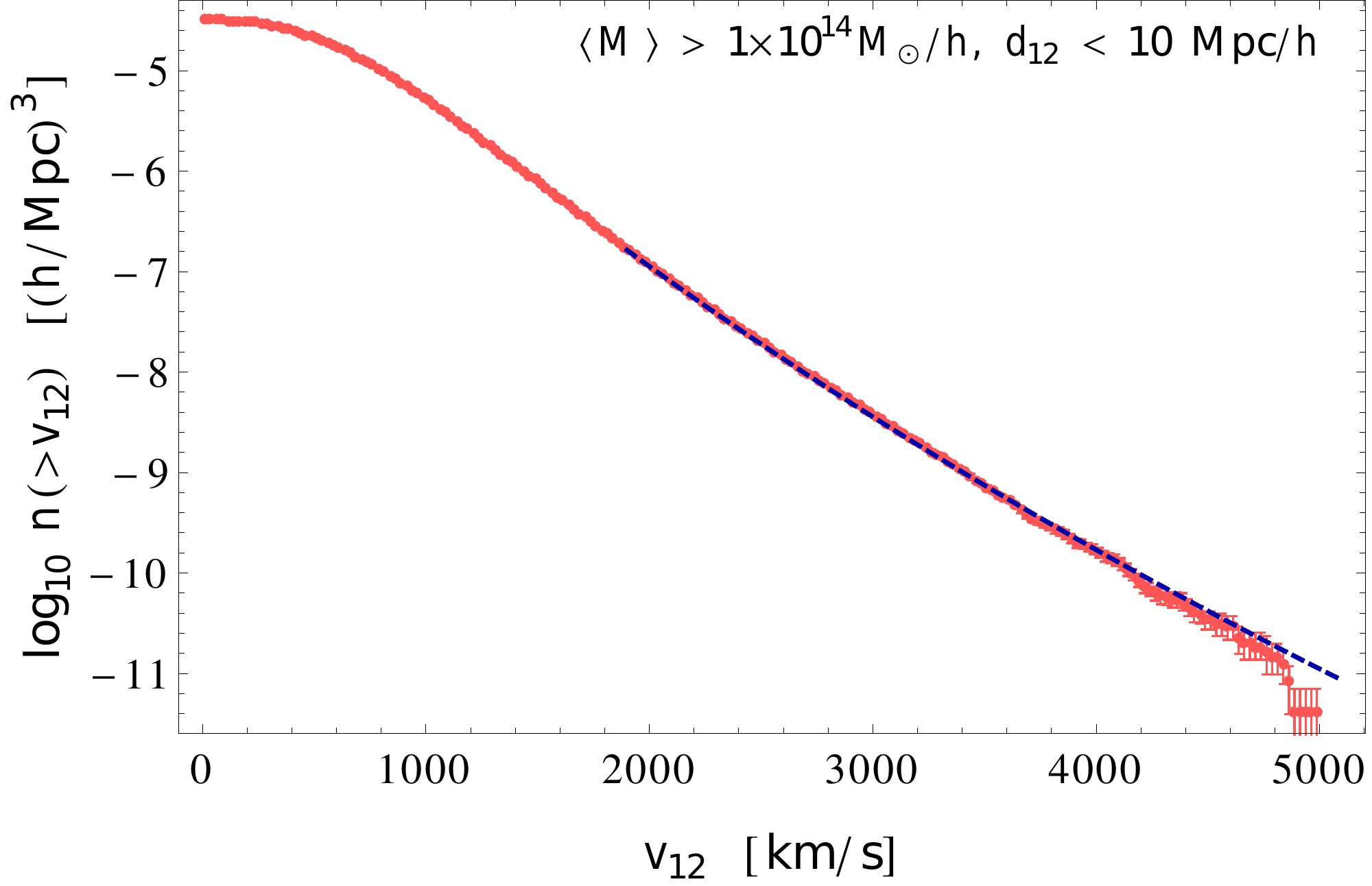}
\caption{Cumulative pairwise velocity distribution from the simulation
  at redshift $z=0$ compared to the best fit Generalised Pareto
  Distribution.}
\label{nv12}
\end{figure}

\begin{figure}
\centering
 \includegraphics[width=0.5\columnwidth]{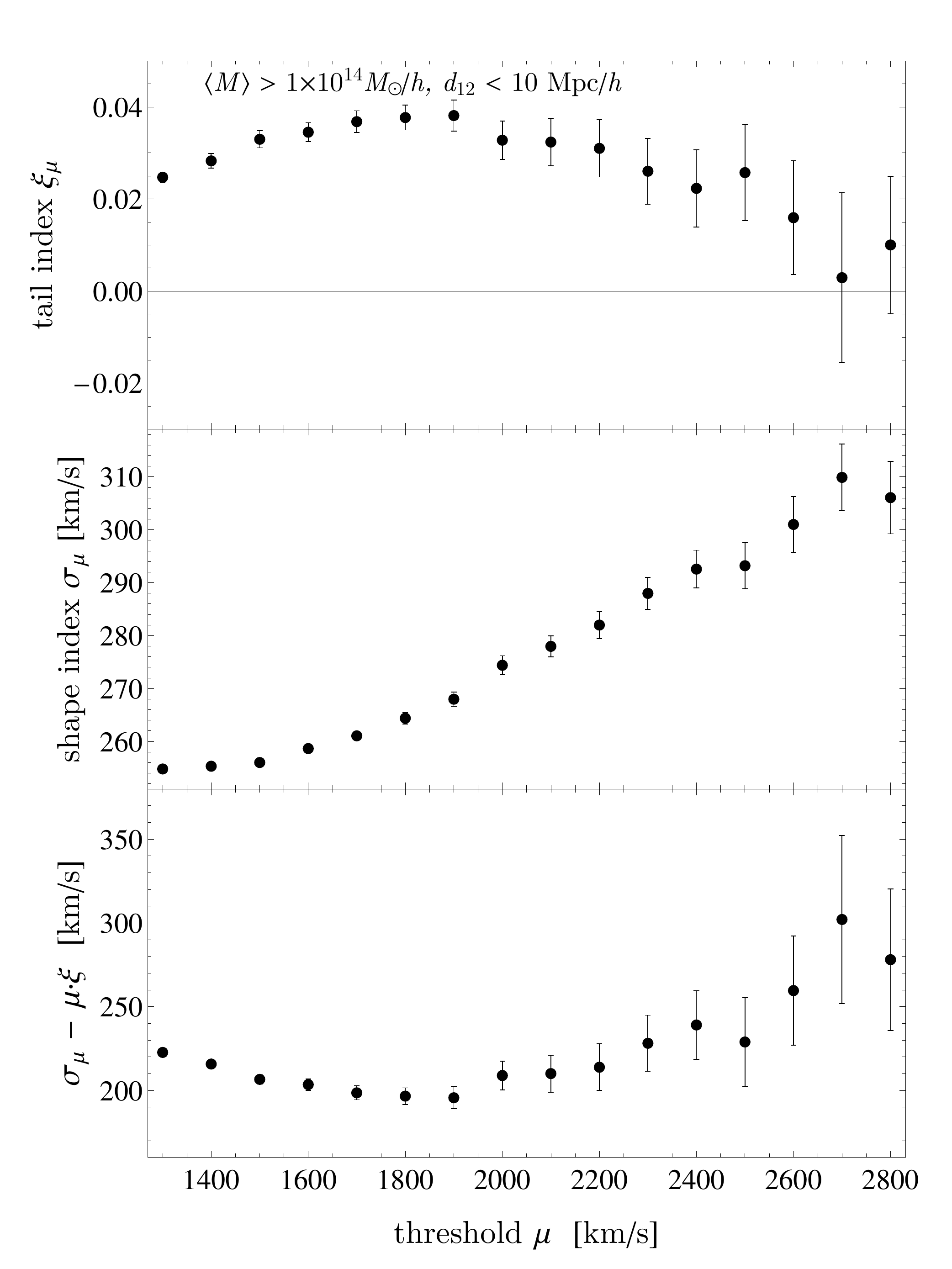}
\caption{Stability plots for the EVS tail fitting exercise. The most
  appropriate threshold $\mu$ is seen to be around 1900\,km/s where
  $\xi_\mu$ is approximately constant and $\sigma_{\mu}$ increases
  linearly.}
\label{stability}
\end{figure}

We have shown above that the number of Bullet-like systems has a
non-trivial dependence on the collisional parameters, which are
moreover \emph{correlated} with each other. Therefore, the expected
number of Bullet-like systems depends strongly on the adopted
definition of such a system. A conservative (i.e. \emph{over}-)
estimate of the number of Bullet-like systems within a cosmic volume
up to some redshift can be obtained by choosing cuts in the
collisional parameters that are less extreme than those characterising
1E\,0657-56. Accordingly, we adopt the following conditions on the
average mass, separation, and the relative velocity of the halo pairs:
$\langle M \rangle > 10^{14} M_{\odot}/h,\text{ } d_{12}\leq
10\,\text{Mpc/h}, \text{and } v_{12} >3000\,\text{km/s}$.
This is comparable to the cuts made in
Refs.\cite{bouillot_probing_2014} and
\cite{thompson_pairwise_2012}. Any additional cuts in the separation
and the angle reduce the number of Bullet-like objects, thus
\emph{exacerbating} any tension of $\Lambda$CDM with observations. The
pairwise velocity distribution and its cumulative version are shown in
Figs.~\ref{dndv12} and \ref{nv12}, where the errorbars have been
estimated by a bootstrap technique. We fit the tail to the GPD form
using the maximum likelihood method. The stability analysis is
presented in Fig.~\ref{stability}. The appropriate choice for the
threshold $\mu$ is around 1900\,km/s; below this the events in the
distribution are normal while above this threshold the variance
increases substantially and the bias due to the finite simulation box
appears (i.e. very rare events are missing altogether).  Therefore the
tail of the pairwise velocity distribution is well characterised by:
$\mu \simeq1900$\,km/s, $\xi = 0.038 \pm 0.003$ and $\sigma = 268.0
\pm 1.4$\,km/s. This is broadly consistent with the results of
Ref.\cite{bouillot_probing_2014}. Thus the extreme events in the tail
of the pairwise velocity distribution are \emph{not} drawn from a
Gaussian-like distribution ($\xi=0$) as has been assumed previously
\citep{lee_bullet_2010,thompson_pairwise_2012,thompson_rise_2014}.

We calculate the expected number of Bullet-like systems as defined
above ($\langle M \rangle > 10^{14} M_{\odot}/h, d_{12} \leq
10\,\text{Mpc}/h, \text{and } v_{12} >3000\,\text{km/s}$) up to
$z=0.3$ (where 1E\,0657-56 is located) and $z=0.5$ (where the initial
conditions for the collision are known). The corresponding effective
volumes from Eq.~(\ref{V_eff}) are $V_{\rm eff}(<z=0.3) \simeq
4.6$\,(Gpc/$h)^3$ and $V_{\rm eff}(<z=0.5) \simeq
13$\,(Gpc/$h)^3$. Using this and the cumulative pairwise velocity
distribution, the expected number of Bullet-like systems is:
\begin{equation}
N(<z=0.3)\simeq17^{+6}_{-5}, \text{ and } N(<z=0.5) \simeq 47^{+8}_{-7},
\end{equation}
where the variance has been estimated by sampling the subvolumes of
size $V_{\rm eff}$ from the full N-Body simulation. 

We now focus on the expected number of objects as or more extreme than
1E\,0657-56 in particular. Making cuts in the collisional parameters
similar to Ref.\cite{thompson_rise_2014}:
\begin{align}
\langle M \rangle > (M_{\rm main}+M_{\rm Bullet})/2 &=5.67\times
10^{14} M_{\odot}/h, \quad &M_{1,2} > 10^{14}M_{\odot}/h, \nonumber
\\ d_{12} &\leq 10\,\text{Mpc}/h, \quad &v_{12} > 3000\,\text{km/s,}
\label{thomcuts} 
\end{align}
we obtain:
\begin{equation}
N(<z=0.3)\simeq1.3^{+2.0}_{-0.6} \quad \text{ and } \quad
N(<z=0.5)\simeq2.5^{+2.1}_{-1.2},
\end{equation}
However since 1E\,0657-56 is observed shortly \emph{after} the
collision we should require further that the halo pairs must be moving
away from each other, i.e.  $\cos \theta >0$. This leads to:
\begin{equation}
N(<z=0.3)\simeq0.1 \quad \text{  and  } \quad N(<z=0.5)\simeq0.15
\end{equation}
Since the pairwise velocity distribution is steeply descending,
increasing the relative velocity $v_{12}$ up to the 4500 km/s velocity
of the shock front in the 1E\,0657-56 merger would \emph{decrease}
this number further by two orders of magnitude, as we see from
Fig.~\ref{masscuts}.

About a dozen other merging clusters have been observed, each with a
different set of collisional parameters. Since a cluster collision is
expected to take a short time compared to the cosmic time we can
consider events both before and after collision by setting $\cos
\theta <-0.9$ or $\cos \theta >0.9$. Requiring $v_{12}\geq 4000$\,km/s
in addition to the mass cuts in Eq.(\ref{thomcuts}), we find only 4
halo pairs in the full (Hubble) volume of the simulation (see Table
\ref{table}).
\begin{table}
\centering
\begin{tabular}{|c|c|c|c|c|}
\hline $M_{1} \left[ M_{\odot}/h \right] $ & $M_{2} \left[ M_{\odot}/h
  \right]$ & $d_{12} \left[\text{Mpc/h}\right]$ & $v_{12}\left[
  \text{km/s} \right]$ & $\cos \theta$ \\ \hline $1.73\times 10^{15}$
& $1.1\times 10^{14}$ & $6.9$ & $4262$ & $-0.99$ \\ $3.11\times
10^{15}$ & $1.2\times 10^{14}$ & $8.3$ & $4140$ & $-0.96$
\\ $3.0\times 10^{15}$ & $1.3\times 10^{14}$ & $2.7$ & $4121$ &
$-0.93$ \\ $3.8\times 10^{15}$ & $1.4\times 10^{14}$ & $3.5$ & $4132$
& $-0.94$ \\ \hline
\end{tabular}
\caption{Colliding halo pairs with the mass cuts from
  Eq.(\ref{thomcuts}) and $v_{12}>4000$\,km/s in the (8\,Gpc/$h)^3$
  Dark Sky simulation. All collisions are selected to be head-on
  ($|\cos \theta| >0.9$). }
\label{table}
\end{table}
Hence, the expected number of such systems is $\langle N
\rangle(<z=0.3) \simeq 0.02$, leading to the probability, $p(N \geq
1)=1-{\rm Poisson} (N=0 \mid \langle N \rangle=0.02) \simeq 0.02$, of
having at least one such system in a cosmic volume up to redshift
$z=0.3$. Furthermore, setting $v_{12}\geq 4500$\,km/s we find
\emph{no} candidate halo pairs. This places an upper limit of 0.005 on
the probability $p(N \geq 1)$ of having at least one system with such
an extreme relative velocity up to $z=0.3$.

For observers, an approximate formula for the number of colliding
clusters expected up to a specified redshift, given specific
collisional parameters, might be of interest:
\begin{align}
&N(<z; ~<d_{12}; ~>v_{12}; ~>\langle M \rangle ; ~>\cos\theta) \approx
  A \langle M \rangle ^ a d_{12}^b z^{c} (1-\cos\theta)^{d} \times \nonumber \\ 
&\exp\left[ \alpha \times (
  d_{12}/\langle M \rangle) + \cos^{2}\theta (\beta +\gamma \times
  \langle M \rangle + \delta \times v_{12}) + v_{12}(\epsilon + \zeta
  \times d_{12} ) + \eta\times \langle M \rangle z \right],
  \label{number}
\end{align}
where $(A,a,b,c,d,\alpha,\beta,\gamma,\delta,\epsilon,\zeta,\eta) =
(80.7 \times 10^{3}, 0.26, 0.93, 2.78, 1.44, 0.22, 1.15, -0.071, 2.50
\times 10^{-4}, -3.43 \times 10^{-3}, -4.82 \times 10^{-5},
-0.58)$. Our fit is valid to within 10\% for $10^{14} M_{\odot}/h
\lesssim \langle M \rangle \lesssim 7\times 10^{14} M_{\odot}/h$, $z <
0.6$, $3 \mathrm{\,Mpc/}h \lesssim d_{12} \lesssim 10
\mathrm{\,Mpc/}h$, $\cos \theta \lesssim 0.9$, and $2000
\mathrm{\,km/s} \lesssim v_{12} \lesssim 4000
\mathrm{\,km/s}$. However it becomes unreliable at higher velocities
and masses, as well as at lower separations. The effective volumes
(\ref{V_eff}) used in the expression (\ref{number}) above are in fact
estimated from a set of smaller simulations
\cite{rasera_introducing_2010}.
 
\section{Conclusions}

We have studied the prevalence of rare DM halo collisions in
$\Lambda$CDM cosmology using the pairwise velocity distribution for
halos extracted from a N-body simulation with volume comparable to the
observable universe and the finest resolution to date. Our approach
differs from previous studies that attempt to quantify the probability
that a cluster and its subcluster, given some masses and separation,
will have a relative velocity as high as 1E\,0657-56. We find that
such a definition of probability can lead to paradoxical conclusions,
so instead we investigate the dependence of the expected number of
Bullet-like systems on the collisional parameters, as well as the
correlations among them. We demonstrate that the expected number of
halo pairs is very sensitive to cuts in the parameters defining the
mergers. Given recent observations of more merging clusters, we
provide a formula for the expected number of halo-halo collisions with
specified collisional parameters up to some redshift.

The tail of the pairwise velocity distribution for the colliding halos
is modelled using Extreme Values Statistics to demonstrate that it is
fatter than a Gaussian. Hence, the combination of a
configuration-space based halo finder, the assumption of a
Gaussian-like tail, small simulation boxes, and poor simulation
resolutions, have resulted in \emph{underestimation} of the number of
high-velocity mergers in previous studies.

We find that only about 0.1 systems like the Bullet Cluster
1E\,0657-56 (where the collision has occurred \emph{already}) can be
expected up to $z=0.3$. Increasing the relative velocity to 4500\,km/s
--- the shock front velocity deduced from X-ray observations of
1E\,0657-56 --- \emph{no} candidate systems are found in the
simulation. Thus the existence of 1E\,0657-56 is only marginally
compatible with the $\Lambda$CDM cosmology, provided the relative
velocity of the two colliding clusters is indeed as low as suggested
by hydrodynamical simulations. Hence if more such systems are found
this would challenge the standard cosmological model.

\section{Acknowledgements}

DK thanks STFC for support and SS acknowledges a DNRF Niels Bohr
Professorship. We are grateful to the anonymous Referee for
emphasising the importance of the fact that the two clusters in
1E\,0657-56 have collided already, and for several other helpful
suggestions.

\bibliographystyle{JHEP.bst}
\bibliography{refs_balt}

\providecommand{\href}[2]{#2}\begingroup\raggedright\begin{thebibliography}{10}

\bibitem{clowe_weak-lensing_2004}
D.~{Clowe}, A.~{Gonzalez}, and M.~{Markevitch}, {\it {Weak-Lensing Mass
  Reconstruction of the Interacting Cluster 1E 0657-558: Direct Evidence for
  the Existence of Dark Matter}},  {\em Astrophys. J.} {\bf 604} (Apr., 2004)
  596--603, [\href{http://arxiv.org/abs/astro-ph/0312273}{{\tt
  astro-ph/0312273}}].

\bibitem{bradac_strong_2006}
M.~{Brada{\v c}}, D.~{Clowe}, A.~H. {Gonzalez}, P.~{Marshall}, W.~{Forman},
  C.~{Jones}, M.~{Markevitch}, S.~{Randall}, T.~{Schrabback}, and
  D.~{Zaritsky}, {\it {Strong and Weak Lensing United. III. Measuring the Mass
  Distribution of the Merging Galaxy Cluster 1ES 0657-558}},  {\em Astrophys.
  J.} {\bf 652} (Dec., 2006) 937--947,
  [\href{http://arxiv.org/abs/astro-ph/0608408}{{\tt astro-ph/0608408}}].

\bibitem{markevitch_direct_2004}
M.~{Markevitch}, A.~H. {Gonzalez}, D.~{Clowe}, A.~{Vikhlinin}, W.~{Forman},
  C.~{Jones}, S.~{Murray}, and W.~{Tucker}, {\it {Direct Constraints on the
  Dark Matter Self-Interaction Cross Section from the Merging Galaxy Cluster 1E
  0657-56}},  {\em Astrophys. J.} {\bf 606} (May, 2004) 819--824,
  [\href{http://arxiv.org/abs/astro-ph/0309303}{{\tt astro-ph/0309303}}].

\bibitem{farrar_new_2007}
G.~R. {Farrar} and R.~A. {Rosen}, {\it {A New Force in the Dark Sector?}},
  {\em Phys. Rev. Lett.} {\bf 98} (Apr., 2007) 171302,
  [\href{http://arxiv.org/abs/astro-ph/0610298}{{\tt astro-ph/0610298}}].

\bibitem{lage_constrained_2014}
C.~{Lage} and G.~{Farrar}, {\it {Constrained Simulation of the Bullet
  Cluster}},  {\em Astrophys. J.} {\bf 787} (June, 2014) 144,
  [\href{http://arxiv.org/abs/1312.0959}{{\tt arXiv:1312.0959}}].

\bibitem{lage_bullet_2014}
C.~{Lage} and G.~R. {Farrar}, {\it {The Bullet Cluster is not a Cosmological
  Anomaly}},  {\em ArXiv e-prints} (June, 2014)
  [\href{http://arxiv.org/abs/1406.6703}{{\tt arXiv:1406.6703}}].

\bibitem{mastropietro_simulating_2008}
C.~{Mastropietro} and A.~{Burkert}, {\it {Simulating the Bullet Cluster}},
  {\em Mon. Not. R. Astron. Soc.} {\bf 389} (Sept., 2008) 967--988,
  [\href{http://arxiv.org/abs/0711.0967}{{\tt arXiv:0711.0967}}].

\bibitem{springel_speed_2007}
V.~{Springel} and G.~R. {Farrar}, {\it {The speed of the `bullet' in the
  merging galaxy cluster 1E0657-56}},  {\em Mon. Not. R. Astron. Soc.} {\bf
  380} (Sept., 2007) 911--925,
  [\href{http://arxiv.org/abs/astro-ph/0703232}{{\tt astro-ph/0703232}}].

\bibitem{milosavljevic_cluster_2007}
M.~{Milosavljevi{\'c}}, J.~{Koda}, D.~{Nagai}, E.~{Nakar}, and P.~R. {Shapiro},
  {\it {The Cluster-Merger Shock in 1E 0657-56: Faster than a Speeding
  Bullet?}},  {\em Astron. J. Lett.} {\bf 661} (June, 2007) L131--L134,
  [\href{http://arxiv.org/abs/astro-ph/0703199}{{\tt astro-ph/0703199}}].

\bibitem{thompson_pairwise_2012}
R.~{Thompson} and K.~{Nagamine}, {\it {Pairwise velocities of dark matter
  haloes: a test for the {$\Lambda$} cold dark matter model using the bullet
  cluster}},  {\em Mon. Not. R. Astron. Soc.} {\bf 419} (Feb., 2012)
  3560--3570, [\href{http://arxiv.org/abs/1107.4645}{{\tt arXiv:1107.4645}}].

\bibitem{hayashi_how_2006}
E.~{Hayashi} and S.~D.~M. {White}, {\it {How rare is the bullet cluster?}},
  {\em Mon. Not. R. Astron. Soc.} {\bf 370} (July, 2006) L38--L41,
  [\href{http://arxiv.org/abs/astro-ph/0604443}{{\tt astro-ph/0604443}}].

\bibitem{lee_bullet_2010}
J.~{Lee} and E.~{Komatsu}, {\it {Bullet Cluster: A Challenge to {$\Lambda$}CDM
  Cosmology}},  {\em Astrophys. J.} {\bf 718} (July, 2010) 60--65,
  [\href{http://arxiv.org/abs/1003.0939}{{\tt arXiv:1003.0939}}].

\bibitem{bouillot_probing_2014}
V.~R. {Bouillot}, J.-M. {Alimi}, P.-S. {Corasaniti}, and Y.~{Rasera}, {\it
  {Probing dark energy models with extreme pairwise velocities of galaxy
  clusters from the DEUS-FUR simulations}},  {\em ArXiv e-prints} (May, 2014)
  [\href{http://arxiv.org/abs/1405.6679}{{\tt arXiv:1405.6679}}].

\bibitem{thompson_rise_2014}
R.~{Thompson}, R.~{Dav{\'e}}, and K.~{Nagamine}, {\it {The rise and fall of a
  challenger: the Bullet Cluster in $\Lambda$ Cold Dark Matter simulations}},
  {\em ArXiv e-prints} (Oct., 2014) [\href{http://arxiv.org/abs/1410.7438}{{\tt
  arXiv:1410.7438}}].

\bibitem{behroozi_rockstar_2013}
P.~S. {Behroozi}, R.~H. {Wechsler}, and H.-Y. {Wu}, {\it {The ROCKSTAR
  Phase-space Temporal Halo Finder and the Velocity Offsets of Cluster Cores}},
   {\em Astrophys. J.} {\bf 762} (Jan., 2013) 109,
  [\href{http://arxiv.org/abs/1110.4372}{{\tt arXiv:1110.4372}}].

\bibitem{watson_statistics_2014}
W.~A. {Watson}, I.~T. {Iliev}, J.~M. {Diego}, S.~{Gottl{\"o}ber}, A.~{Knebe},
  E.~{Mart{\'{\i}}nez-Gonz{\'a}lez}, and G.~{Yepes}, {\it {Statistics of
  extreme objects in the Juropa Hubble Volume simulation}},  {\em Mon. Not. R.
  Astron. Soc.} {\bf 437} (Feb., 2014) 3776--3786,
  [\href{http://arxiv.org/abs/1305.1976}{{\tt arXiv:1305.1976}}].

\bibitem{forero-romero_bullet_2010}
J.~E. {Forero-Romero}, S.~{Gottl{\"o}ber}, and G.~{Yepes}, {\it {Bullet
  Clusters in the MARENOSTRUM Universe}},  {\em Astrophys. J.} {\bf 725} (Dec.,
  2010) 598--604, [\href{http://arxiv.org/abs/1007.3902}{{\tt
  arXiv:1007.3902}}].

\bibitem{baldi}
J.~{Lee} and M.~{Baldi}, {\it {Can Coupled Dark Energy Speed up the Bullet
  Cluster?}},  {\em Astrophys. J.} {\bf 747} (Mar., 2012) 45,
  [\href{http://arxiv.org/abs/1110.0015}{{\tt arXiv:1110.0015}}].

\bibitem{skillman_dark_2014}
S.~W. {Skillman}, M.~S. {Warren}, M.~J. {Turk}, R.~H. {Wechsler}, D.~E. {Holz},
  and P.~M. {Sutter}, {\it {Dark Sky Simulations: Early Data Release}},  {\em
  ArXiv e-prints} (July, 2014) [\href{http://arxiv.org/abs/1407.2600}{{\tt
  arXiv:1407.2600}}].

\bibitem{navarro_structure_1996}
J.~F. {Navarro}, C.~S. {Frenk}, and S.~D.~M. {White}, {\it {The Structure of
  Cold Dark Matter Halos}},  {\em Astrophys. J.} {\bf 462} (May, 1996) 563,
  [\href{http://arxiv.org/abs/astro-ph/9508025}{{\tt astro-ph/9508025}}].

\bibitem{hernquist_n-body_1993}
L.~{Hernquist}, {\it {N-body realizations of compound galaxies}},  {\em
  Astrophys. J. Suppl.} {\bf 86} (June, 1993) 389--400.

\bibitem{coles2001introduction}
S.~Coles, {\em An Introduction to Statistical Modeling of Extreme Values}.
\newblock Lecture Notes in Control and Information Sciences. Springer, 2001.

\bibitem{rasera_introducing_2010}
Y.~Rasera, J.-M. Alimi, J.~Courtin, F.~Roy, P.-S. Corasaniti, A.~Füzfa, and
  V.~Boucher, {\it Introducing the dark energy universe simulation series
  ({DEUSS})},  vol.~1241, pp.~1134--1139, June, 2010.
\newblock \href{http://arxiv.org/abs/1002.4950}{{\tt arXiv:1002.4950}}.

\bibitem{tinker_large-scale_2010}
J.~L. {Tinker}, B.~E. {Robertson}, A.~V. {Kravtsov}, A.~{Klypin}, M.~S.
  {Warren}, G.~{Yepes}, and S.~{Gottl{\"o}ber}, {\it {The Large-scale Bias of
  Dark Matter Halos: Numerical Calibration and Model Tests}},  {\em Astrophys.
  J.} {\bf 724} (Dec., 2010) 878--886,
  [\href{http://arxiv.org/abs/1001.3162}{{\tt arXiv:1001.3162}}].

\bibitem{tinker_mass--light_2005}
J.~L. {Tinker}, D.~H. {Weinberg}, Z.~{Zheng}, and I.~{Zehavi}, {\it {On the
  Mass-to-Light Ratio of Large-Scale Structure}},  {\em Astrophys. J.} {\bf
  631} (Sept., 2005) 41--58, [\href{http://arxiv.org/abs/astro-ph/0411777}{{\tt
  astro-ph/0411777}}].

\end{thebibliography}\endgroup

\end{document}